\documentclass[12pt,a4paper]{article}
\usepackage{epsfig,graphics}
\begin{document}

\author{G. F. Bressange\thanks{E-mail : georges.bressange@ucd.ie}\\
\small Mathematical Physics Department\\
\small  National University of Ireland Dublin, Belfield, Dublin 4, Ireland}

\title{On the extension of the concept of thin shells to the
  Einstein-Cartan theory}
\date{}
\maketitle

\begin{abstract}
This paper develops a theory of thin shells within the context of
the Einstein-Cartan theory by extending the known formalism of
general relativity. In order to perform such an extension, we
require the general non symmetric stress-energy tensor to be
conserved leading, as Cartan pointed out himself, to a strong
constraint relating curvature and torsion of spacetime. When we
restrict ourselves to the class of space-times satisfying this
constraint, we are able to properly describe thin shells and
derive the general expression of surface stress-energy tensor
both in its four-dimensional and in its three-dimensional
intrinsic form. We finally derive a general family of static
solutions of the Einstein-Cartan theory exhibiting a natural
family of null hypersurfaces and use it to apply our formalism to
the construction of a null shell of matter.

\end{abstract}

\thispagestyle{empty}
\newpage


\section{Introduction}\indent
The purpose of this paper is to extend the notion of singular hypersurface
existing in general relativity to the context of the
Einstein-Cartan theory in a coherent way in order to include the
effects of torsion originating in the spinning properties of the matter
distribution. The role and the importance of such objects
has been extensively studied within the context of the Einstein
theory of gravitation and we refer the reader to existing reviews
like \cite{BI} and the references contained therein for their various
aspects. A singular hypersurface is a geometric
representation in spacetime of the interface existing between
two coexisting phases of matter such that the curvature tensor
suffers a discontinuity of the Dirac type at the inter-phase. This results in
a two-dimensional concentrated distribution of matter whose history
in spacetime corresponds to this hypersurface, a distribution of matter
which is usually referred to as a thin shell or surface layer.
Singular hypersurfaces stand for the
idealization of a vast range of physical situations such as for
example the false-vacuum bubbles used in inflationary cosmology (see
the introduction of \cite{BI} and the references listed) as well as in
extended inflation in Brans-Dicke models (see \cite{BB1} for a review).
They also have been successfully used to describe
impulsive gravitational waves coexisting with null shells of matter by
recently providing a model of supernovae \cite{BBH1} in which the
burst of gravitational radiation and the burst of neutrinos are
respectively modeled by a gravitational impulsive wave and a null
shell of matter such that the wave-fronts of the wave and the shell
share a singular null hypersurface as their common history. More
recently, the formalism of singular hypersurfaces derived in
\cite{BI}\cite{BBH1}
has been successfully applied to the description of the peeling
properties of a light-like signal propagating through a general
Bondi-Sachs model
\cite{BrH1}.

   The Einstein-Cartan theory of gravitation was originally suggested
   by Elie Cartan in 1922 \cite{Cartan1} as a natural generalisation
   of Einstein's theory of general relativity by extending the model
   of spacetime into a four-dimensional manifold endowed with a linear
   connection still compatible with the metric tensor but non
   symmetric. As a result, the torsion tensor, defined by Cartan as
   the antisymmetric part of the connection, plays a non trivial role
   in the geometry and Cartan's idea was to relate this torsion to the
   density of intrinsic angular momentum \cite{Cartan2}\cite{Cartan}.
   Cartan's ideas reemerged in the 60's with the works of Sciama \cite{Sciama}
   and Kibble \cite{Kibble} who independently developed a gauge theory
   approach to gravitation from which the modern $U_4$ theory
   emerged. For this reason, the Einstein-Cartan theory is also known
   as the Einstein-Cartan-Sciama-Kibble theory. Later, it was proved
   that this theory of gravitation could lead to non singular
   cosmological models \cite{Kopc}. The mathematical structure of
   the Einstein-Cartan equations was also analyzed into a new
   perspective by A. Trautman \cite{Traut}, using tensor-valued
   differential forms, who proved that the Cartan equations determine
   the linear connection only up to projective transformations and
   that the condition for the connection to be metric-compatible is
   exactly the condition which removed this arbitrarness. For a more
   detailed historical development, we refer to the section I.6 of
   \cite{Hehl}.

   It is therefore of great interest to investigate the problem of
   discontinuities within this theory by analysing the matching
   conditions the various fields have to satisfy across a given
   singular hypersurface whose definition will be exposed within this
   paper. Gravitational matching conditions have first been considered
   in the Einstein-Cartan theory by Arkuszewski, Kopczy\'nski and
   Pomomariev \cite{AKP} who derived the
   matching and the junction conditions for surfaces
   characterized by the continuity of the second fundamental form,
   also called extrinsic curvature.
   This corresponds to boundary surfaces in general relativity where the first
   derivatives of the metric tensor are continuous across the
   hypersurface, such as for example the surface separating a star from
   the surrounding vacuum. But this reference didn't consider the case of more
   irregular discontinuities arising when the metric tensor is assumed
   to be only continuous across the surface with a jump in the
   extrinsic curvature which corresponds to thin shells. Moreover,
   they considered only timelike
   hypersurfaces. This paper is an attempt to describe these more
   general discontinuities by providing a formalism to derive the
   physical content of thin shells of any type, in particular
   null. This will be achieved by extending the Barrab\`es-Israel
   formalism of general relativity \cite{BI} to the Einstein-Cartan theory in
   order to include the effects of spin.

   The paper is organized as follows. In section \ref{II}, we present the
   Einstein-Cartan theory and summarize all the useful equations and
   identities we will need in this paper. We then discuss the type
   of spacetimes we will restrict to
   in order to derive our shell formalism. The definition of a
   singular hypersurface and the shell formalism are
   presented in section \ref{III}. The stress-energy tensor of the
   shell is derived both in its four dimensional and
   intrinsic versions. Finally, in section \ref{IV}, we construct a
   general family of static Einstein-Cartan solutions
   with non-trivial torsion which contains a natural family of null
   hypersurfaces and we use it to construct a null shell of matter. The
   surface energy density of the shell is characterized by the jump
   in the local mass function of the two surrounding spacetimes and the
   surface pressure is given in terms of the jump in the components of the
   torsion across the hypersurface.


\section{The Einstein-Cartan theory}\label{II}

        The Einstein-Cartan theory considers a four-dimensional
manifold ${\cal M}$ as a model of spacetime, endowed with a
metric tensor ${\bf g}$ of signature $(-,+,+,+)$ satisfying
the same regularity conditions as in General Relativity (three
times continuously differentiable) but characterized by a non
symmetric connection. The components of this connection are denoted
$\Gamma^{\lambda}_{\ \mu\nu}$ in a local coordinate system
$\{x^{\mu}\}$, Greek indices taking the values 0,1, 2, 3.
This implies the existence of a non-vanishing torsion tensor ${\bf Q}$
such that
\begin{equation}
Q^{\lambda}_{\ \mu\nu}=\Gamma^{\lambda}_{\ \nu\mu}
                       -\Gamma^{\lambda}_{\ \mu\nu}\ .
\end{equation}
The metric tensor $g_{\mu\nu}$ is still assumed to be
metric-compatible with the covariant derivative associated with the
above connection i.e. $\nabla_{\mu}g_{\nu\lambda}\,=\,0$. It is useful
to display the existing relation between the Einstein-Cartan metric
compatible linear connection and the Riemannian connection constructed
out of the metric $g_{\mu\nu}$ represented by its Christoffel symbols
${^R}\Gamma^{\lambda}_{\ \mu\nu}$, namely
\begin{equation}\label{RelCon}
        \Gamma^{\lambda}_{\ \mu\nu}\,=\,{^R}\Gamma^{\lambda}_{\ \mu\nu}
                                +\chi^{\lambda}_{\ \mu\nu}\hspace{0.7cm},
\end{equation}
with
\begin{equation}
        {^R}\Gamma^{\lambda}_{\ \mu\nu}\,=\,{1\over 2}\,g^{\lambda\rho}\,
        (\partial_{\mu}g_{\nu\rho}+\partial_{\nu}g_{\mu\rho}
        -\partial_{\rho}g_{\mu\nu})\hspace{0.7cm},
\end{equation}
where we introduced
\begin{equation}\label{contortion}
        \chi_{\lambda\mu\nu}\,=\,-{1\over 2}\,\left(Q_{\lambda\mu\nu}
                                +Q_{\mu\nu\lambda}+Q_{\nu\mu\lambda} \right)\,
                                =\, -\chi_{\mu\lambda\nu}\hspace{0.7cm},
\end{equation}
which is often referred as the ``contortion''or ``defect'' of the
connection (see references \cite{Hehl} and \cite{AKP}).

In the local coordinate system $\{x^{\mu}\}$, the components of the
curvature tensor have the familiar expression
\begin{equation}\label{Curv}
R^{\mu}_{\ \nu\lambda\rho}=\partial_{\lambda}\Gamma^{\mu}_{\ \nu\rho}
                                -\partial_{\rho}\Gamma^{\mu}_{\ \nu\lambda}
                +\Gamma^{\mu}_{\ \kappa\lambda}\Gamma^{\kappa}_{\ \nu\rho}
                -\Gamma^{\mu}_{\ \kappa\rho}\Gamma^{\kappa}_{\
                  \nu\lambda} \hspace{0.7cm}.
\end{equation}

        The symmetries of the Riemann tensor which remain within this
        context are
\begin{equation}\label{sym1}
R_{\mu\nu\lambda\rho}\,=\,-R_{\mu\nu\rho\lambda}\hspace{0.7cm},
\end{equation}
and
\begin{equation}\label{sym2}
R_{\mu\nu\lambda\rho}\,=\,-R_{\nu\mu\lambda\rho}\hspace{0.7cm}.
\end{equation}
The symmetry (\ref{sym1}) comes from the general definition of a
curvature operator and the symmetry (\ref{sym2}) follows from the fact
that the connection is metric compatible.

        The cyclic identity is replaced by
\begin{equation}\label{cyclic}
       R^{\mu}_{\ [\nu\lambda\rho]}\,=\,\nabla_{[\nu}Q^{\mu}_{\ \lambda\rho]}
                         +Q^{\mu}_{\ \kappa[\nu}Q^{\kappa}_{\
                         \lambda\rho]} \hspace{0.7cm},
\end{equation}
which is also called Bianchi identity for the torsion.

        The property of symmetry by interchanging pairs of indices in
        the Riemann tensor is no longer true in this non-Riemannian
        spacetime but is replaced by the following identity, obtained
        using four times the identity (\ref{cyclic}):
\begin{equation}\label{sym3}
R_{\mu\nu\lambda\rho}\,=\,R_{\lambda\rho\mu\nu}+S^{(1)}_{\mu\nu\lambda\rho}
                                        +S^{(2)}_{\mu\nu\lambda\rho}
\hspace{0.7cm},
\end{equation}
with
\begin{equation}\label{S1}
           S^{(1)}_{\mu\nu\lambda\rho} \,=\,\nabla_{\nu}\chi_{\lambda\rho\mu}
                +\nabla_{\lambda}\chi_{\mu\nu\rho}
                +\nabla_{\rho}\chi_{\nu\mu\lambda}
                +\nabla_{\mu}\chi_{\rho\lambda\nu}\hspace{0.7cm},
\end{equation}
and
\begin{equation}\label{S2}
 S^{(2)}_{\mu\nu\lambda\rho}\,=\,\psi_{\mu\nu\lambda\rho}
              +\psi_{\lambda\mu\rho\nu}
              +\psi_{\rho\mu\nu\lambda}+\psi_{\nu\lambda\mu\rho}
              \hspace{0.7cm},
\end{equation}
where we introduced
\begin{equation}
        \psi_{\mu\nu\lambda\rho}\,=\,{\scriptstyle{1\over 2}}
                \left(Q_{\mu\kappa\nu}Q^{\kappa}_{\ \lambda\rho}
                +Q_{\mu\kappa\lambda}Q^{\kappa}_{\ \rho\nu}
                +Q_{\mu\kappa\rho}Q^{\kappa}_{\ \nu\lambda} \right)
                \,=\,-\psi_{\mu\lambda\nu\rho}\,=\,-\psi_{\mu\rho\lambda\nu}
                \hspace{0.7cm}.
\end{equation}
The Bianchi identity is now
\begin{equation}\label{Bianchi}
           \nabla_{[\lambda}R^{\mu}_{\ \mid\nu\mid\rho\sigma]}
           +R^{\mu}_{\ \nu\kappa[\lambda}Q^{\kappa}_{\
           \rho\sigma]}\,=\,0 \hspace{0.7cm}.
\end{equation}

        The basic field equations of the Einstein-Cartan theory are
\begin{equation}\label{FE1}
G_{\mu\nu}\,=\,8 \pi T_{\mu\nu}\hspace{0.7cm},
\end{equation}
and
\begin{equation}\label{FE2}
        Q^{\mu}_{\ \nu\lambda}+\delta^{\mu}_{\nu}Q^{\rho}_{\ \lambda\rho}
                                -\delta^{\mu}_{\lambda}Q^{\rho}_{\ \nu\rho}
                                        \,=\,8 \pi S^{\mu}_{\
                                          \nu\lambda}
                                        \hspace{0.7cm},
\end{equation}
where $S^{\mu}_{\ \nu\lambda}$ is the spin tensor representing the
density of intrinsic angular momentum, related to the torsion tensor
in a purely algebraic way. $T_{\mu\nu}$ represents the stress-energy
tensor of the spacetime matter content. In the field equation
(\ref{FE1}), $G_{\mu\nu}$ stands for the Einstein tensor constructed
out of the Ricci tensor $R_{\mu\nu}\,=\,R^{\lambda}_{\ \mu\lambda\nu}$
by means of the non symmetric connection $\Gamma^{\lambda}_{\
  \mu\nu}$. We should insist on the fact that this makes the Einstein
tensor non symmetric and therefore, in general, the stress-energy
tensor of the spacetime is a non symmetric quantity within this framework.
When the torsion vanishes so does the spin and these field equations
reduce to the Einstein field equations of general relativity.

        Contraction of the identity (\ref{cyclic}) yields
\begin{equation}
  R_{\nu\rho} \,=\,
               R_{\rho\nu}
       +\nabla_{\mu}\,Q^{\mu}_{\ \nu\rho}+\nabla_{\nu}\,Q^{\mu}_{\ \rho\mu}
       -\nabla_{\rho}\,Q^{\mu}_{\ \nu\mu}
       +Q^{\mu}_{\ \kappa\mu}Q^{\kappa}_{\ \rho\nu}\hspace{0.7cm}.
\end{equation}
Regrouping the covariant derivatives and taking into account the
relation between spin and torsion given in (\ref{FE2}), this equation
leads directly to the generalized evolution equation of the spin tensor
\begin{equation}
        \nabla_{\mu}\,S^{\mu}_{\ \nu\rho}\,=\,2\,T_{[\nu\rho]}
                +2\,S^{\mu}_{\ \kappa\mu}Q^{\kappa}_{\ \nu\rho}\hspace{0.7cm},
\end{equation}
which is therefore not conserved in general. In particular, if the
stress-energy tensor of the spacetime is chosen to be symmetric, this
evolution equation is
\begin{equation}
        \nabla_{\mu}\,S^{\mu}_{\ \nu\rho}\,=\,
                2\,S^{\mu}_{\ \kappa\mu}Q^{\kappa}_{\ \nu\rho}
                \hspace{0.7cm}.
\end{equation}
However, one must keep in mind that the torsion does not propagate
(cf. \cite{Hehl}). Indeed, from the algebraic field equation
(\ref{FE2}), it is clear there cannot exist torsion outside the
spinning matter distribution itself (the vanishing of the one is
equivalent to the vanishing of the other) and therefore the torsion
cannot propagate through vacuum. Outside the spinning matter
distribution, the influence of the spin on the evolution of spacetime
can only be seen on the metric tensor but this effect is of higher
order than the effect of mass on the spacetime metric tensor.
In some exact cosmological models known in the Einstein-Cartan
literature (see for instance \cite{Kopc}), the stress-energy tensor of
the spacetime is chosen to be symmetric and the torsion tensor
trace-free. The spin tensor is in this sense conserved.

         The twice contracted Bianchi identity
(\ref{Bianchi}) (which in general relativity directly leads to the
conservation of the Einstein tensor and consequently to the
conservation of the stress-energy tensor) here results in
\begin{equation}\label{2ContBianchi}
   \nabla_{\nu}\,G^{\nu}_{\ \rho} \,=\,
   R^{\mu}_{\ \lambda}\,Q^{\lambda}_{\ \rho\mu}
   +{1\over 2}\,R^{\mu\nu}_{\ \ \lambda\rho}\,Q^{\lambda}_{\ \mu\nu}
\hspace{0.7cm}.
\end{equation}
This proves, in view of the field equation (\ref{FE1}), that the
stress-energy tensor is not conserved in general. This was remarked by
Cartan himself \cite{Cartan} who arrived at the conclusion that, in
order for the stress-energy momentum to be conserved, a geometrical
constraint relating the torsion {\it and} the curvature has to be
imposed, namely
\begin{equation}\label{STC}
        {\cal C}_{\rho}\,:=\,R^{\mu}_{\ \lambda}\,Q^{\lambda}_{\ \rho\mu}
    +{1\over 2}\,R^{\mu\nu}_{\ \ \lambda\rho}\,Q^{\lambda}_{\
    \mu\nu}\,=\,0
  \hspace{0.7cm}.
\end{equation}
In the following sections, we will restrict ourselves to the class of
spacetimes with torsion and curvature satifying this constraint. The
existence of this constraint will also reveal itself to be necessary in
order to properly describe thin shells within this framework as we
will now investigate.


\section{The Shell formalism}\label{III}

        In this section, we develop a technique to describe thin
        shells of matter in the Einstein-Cartan theory by extending
        the known formalism of general relativity. A thin shell
        corresponds to a distribution of matter ideally assumed to be
        concentrated on a two-dimensional surface. Its history in
        spacetime is therefore geometrically represented by a
        three-dimensional hypersurface which can be either timelike,
        spacelike or lightlike. Such a shell is generally constructed
        using the ``cut and paste'' approach of two spacetimes along a
        hypersurface on which the induced metrics of both spacetimes
        are required to agree. Within the framework of general
        relativity, the Barrab\`es-Israel formalism \cite{BI} provides
        a recipe to obtain a unified description of shells of any type
        by extending the known formalism for timelike and spacelike
        shells. Its main feature is a description of the surface
        properties in terms of the jump of the transverse extrinsic
        curvature (generalizing the usual extrinsic curvature - see
        below for the definition) and thus directly obtained as
        functions of the shell intrinsic coordinates. This makes
        possible the four-dimensional coordinates of each side of the
        layer to be chosen freely and independently. In particular, no
        continuous coordinate system is required and we can choose
        coordinates adapted to the peculiar symmetries of each
        spacetime which greatly simplifies the practical
        calculations. In the following paragraphs, we will extend this
        formalism in the presence of a non symmetric connection.

        Let us consider a hypersurface $\Sigma$ which can be either
        timelike, spacelike or lightlike. We assume that $\Sigma$
        results  from the isometric soldering of two hypersurfaces
        $\Sigma^+$ and $\Sigma^-$ bounding respectively two spacetimes
        $M ^+$ and $M ^-$ -see figure 1. We also assume that $M ^-$
        is endowed with
        a metric tensor, a non symmetric connection and a torsion
        tensor whose components are given respectively by
        $g_{\alpha\beta}^-$, $^{-}\Gamma^{\alpha}_{\ \beta\gamma}$ and
        $^{-}Q^{\alpha}_{\ \beta\gamma}$ with respect to a local
        coordinate system $\{x^{\alpha}_-\}$ and that $M ^+$ is
        endowed with a metric tensor, a non symmetric connection and a
        torsion tensor whose components are given respectively by
        $g_{\alpha\beta}^+$, $^{+}\Gamma^{\alpha}_{\ \beta\gamma}$ and
        $^{+}Q^{\alpha}_{\ \beta\gamma}$ with respect to a local
        coordinate system $\{x^{\alpha}_+\}$. $\Sigma$ can thereby
        be considered as a hypersurface imbedded in the hybrid manifold
        ${\cal M}\,=\,M^- \cup M^+$. The metrics $g_{\alpha\beta}^-$
        and $g_{\alpha\beta}^+$ are at least $C^3$-continuously
        differentiable respectively in $M ^-$ and $M ^+$. The
        connections are at least $C^2$-continuously differentiable
        respectively in $M ^-$ and $M ^+$ (from relation
        (\ref{RelCon}), this is equivalent to requiring the torsion
        tensors to be at least $C^2$-continuously differentiable
        respectively in each domain). If $\{\xi^a\}$ is a set of
        intrinsic coordinates on $\Sigma$ (where the Latin indices
        take the values 1, 2, 3), the vectors $e_{(a)}\,=\,{\partial
        \over \partial \xi^a}$ form a basis of tangent vectors to
        $\Sigma$ at each point of $\Sigma$.

\begin{center}
\begin{picture}(0,0)%
\epsfig{file=bres01.pstex}%
\end{picture}%
\setlength{\unitlength}{3355sp}%
\begingroup\makeatletter\ifx\SetFigFont\undefined%
\gdef\SetFigFont#1#2#3#4#5{%
  \reset@font\fontsize{#1}{#2pt}%
  \fontfamily{#3}\fontseries{#4}\fontshape{#5}%
  \selectfont}%
\fi\endgroup%
\begin{picture}(7209,4447)(526,-5903)
\put(2326,-5836){\makebox(0,0)[lb]{\smash{\SetFigFont{12}{14.4}{\rmdefault}{\bfdefault}{\updefault}$\{\xi^a,\,a=1,2,3\}$}}}
\put(4126,-5836){\makebox(0,0)[lb]{\smash{\SetFigFont{12}{14.4}{\rmdefault}{\bfdefault}{\updefault}$ \longrightarrow \,\mathrm{e}_{(a)}\,=\,\frac{\partial\,\,}{\partial \xi^a} $}}}
\put(4951,-2236){\makebox(0,0)[lb]{\smash{\SetFigFont{12}{14.4}{\rmdefault}{\bfdefault}{\updefault}$\mathrm{e_{\scriptstyle{(a)}}}$}}}
\put(1876,-2986){\makebox(0,0)[lb]{\smash{\SetFigFont{10}{12.0}{\rmdefault}{\bfdefault}{\updefault}$\mathrm{N.\,n\,=\,{\eta}^{-1}\,\neq\,0}$}}}
\put(3226,-1936){\makebox(0,0)[lb]{\smash{\SetFigFont{12}{14.4}{\rmdefault}{\bfdefault}{\updefault}$\mathrm{n}$}}}
\put(4276,-1936){\makebox(0,0)[lb]{\smash{\SetFigFont{12}{14.4}{\rmdefault}{\bfdefault}{\updefault}$\mathrm{N}$}}}
\put(6526,-3136){\makebox(0,0)[lb]{\smash{\SetFigFont{12}{14.4}{\rmdefault}{\bfdefault}{\updefault}$\Sigma\,:=\,\Sigma_+\,\equiv\,\Sigma_-$}}}
\put(526,-1636){\makebox(0,0)[lb]{\smash{\SetFigFont{12}{14.4}{\rmdefault}{\bfdefault}{\updefault}$\mathrm{(M^+,\, g_{\alpha\beta}^+,\,{^+}\Gamma^{\gamma}_{\ \alpha\beta},{^+}Q^{\alpha}_{\ \beta\gamma})}$}}}
\put(3826,-4486){\makebox(0,0)[lb]{\smash{\SetFigFont{12}{14.4}{\rmdefault}{\bfdefault}{\updefault}$\mathrm{(M^-,\, g_{\alpha\beta}^-,{^-}\Gamma^{\gamma}_{\ \alpha\beta}, {^+}Q^{\alpha}_{\ \beta\gamma})}$}}}
\end{picture}

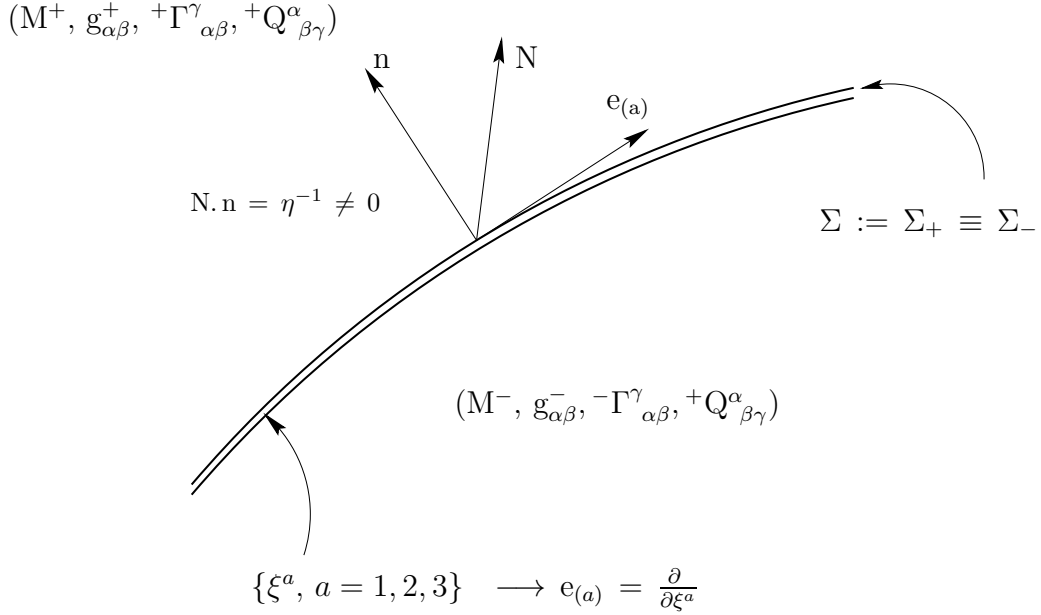
Figure 1 : Isometric soldering of two spacetimes
\end{center}

        For any quantity $F$ regular in ${\cal M} \slash \Sigma$
        (which can be a function aswell as a tensor field) such that
        $F$ tends uniformly to finite left and right limits on
        $\Sigma$, we will denote as usual by $\lbrack F \rbrack$ the
        difference of these limits at a point on $\Sigma$. $\lbrack F
        \rbrack$ is by definition a function defined on $\Sigma$ and
        continuous on $\Sigma$ and is called the jump of $F$ across
        $\Sigma$.

        We will call a singular hypersurface in spacetime $\cal M$
        a hypersurface such that the following regularity conditions
        occur across $\Sigma$ :
\begin{equation}\label{RC1}
        \lbrack g_{\mu\nu} \rbrack \,=\, 0
\end{equation}
\begin{equation}\label{RC2}
        \lbrack e_{(a)}^{\mu}e_{(b)}^{\nu}e_{(c)}^{\lambda}\,
                Q_{\mu\nu\lambda} \rbrack \,=\,\lbrack Q_{abc}\rbrack \,=\, 0
\end{equation}
\begin{equation}\label{RC3}
        \lbrack \Gamma^{\mu}_{\ \nu\lambda} \rbrack \,\neq \, 0
\end{equation}
Condition (\ref{RC1}) expresses the continuity of the metric across
$\Sigma$ which we assumed by construction of $\Sigma$ and this means
that the induced metrics on both sides of the hypersurface are the
same i.e.
\begin{equation}
        g_{\alpha\beta}\,e_{(a)}^{\alpha}e_{(b)}^{\beta}|_{\pm}\,
                =\,g_{ab}\,=\,e_{(a)}\,.\,e_{(b)}\hspace{0.7cm},
\end{equation}
where $g_{ab}$ stands for the the three-dimensional metric intrinsic
to $\Sigma$. Condition (\ref{RC2}) is the requirement that the purely
tangential part (i.e. the projection on $\Sigma$) of the torsion
tensor has to be
continuous across $\Sigma$. As three-dimensional manifold embedded in
an Einstein-Cartan four-dimensional spacetime, $\Sigma$ is also
endowed with a three-dimensional torsion it inherits from the non
symmetric connections of one or other side. Condition (\ref{RC2}) is
therefore the requirement that the induced torsions from both sides
coincide with the intrinsic torsion of $\Sigma$. If we define
$x_{\mu\nu\lambda}\,=\,\lbrack \chi_{\mu\nu\lambda}\rbrack$ the jump
of the defect across the hypersurface and denote by
$x_{abc}\,=\,x_{\mu\nu\lambda}\,e_{(a)}^{\mu}\,e_{(b)}^{\nu}
\,e_{(c)}^{\lambda}$ its projection onto $\Sigma$, we immediately see
from  the definition
(\ref{contortion}) that $x_{abc}\,=\,-{\scriptstyle{1\over2}}\lbrack
Q_{abc}+Q_{bca}+Q_{cba}\rbrack$ and $x_{abc}+x_{bca}\,=\,-\lbrack
Q_{bca}\rbrack$ proving that the vanishing of $\lbrack Q_{abc}\rbrack$
is equivalent to the vanishing of $x_{abc}$. This gives another set of
conditions equivalent to (\ref{RC1})-(\ref{RC3}) by simply replacing
(\ref{RC2}) by $\lbrack \chi_{abc} \rbrack\,=\,0$. We point out
that this is not equivalent to saying that the jump across $\Sigma $ of
the tangential part $S_{abc}$ of the spin tensor vanishes. Indeed,
the field equation (\ref{FE2}) only tells us that $\lbrack
Q_{abc}+Q_{bca}+Q_{cab}\rbrack\,=\,\lbrack
S_{abc}+S_{bca}+S_{cab}\rbrack$ showing that a necessary (but not
sufficient) condition for (\ref{RC2}) is
$\lbrack S_{abc}+S_{bca}+S_{cab}\rbrack\,=\,0$ and {\it not}
$\lbrack S_{abc} \rbrack \,=\,0$.

The condition (\ref{RC3}) means that we assume a jump in the
connection across the hypersurface and it is clear from the relation
(\ref{RelCon}) that this discontinuity results from the
discontinuities of the first derivatives of the metric tensor
and from the discontinuities of the torsion.

         Such a hypersurface, as we will see below, is characterized by
         a jump in the (transverse) extrinsic curvature of the
         hypersurface  and should be distinguish from the boundary
         surfaces  studied in
         reference \cite{AKP} where the extrinsic curvature is assumed
         to be  continuous (this reference, incidently, is confined to
         the timelike case).\footnote{This continuity condition is
         proved to  be equivalent to the regularity of the full
         Riemann tensor, a condition which is not required in the
         context of thin shells.}

        Following the procedure of reference \cite{BI}, we now
        describe how to obtain the expression for the stress-energy
        tensor carried by the singular hypersurface $\Sigma$ in its
        four-dimensional form and in its intrinsic form. We first
        consider a single coordinate system $\{y^{\mu}\}$ of $\cal M$
        maximally smooth across $\Sigma$. Let $n$ be a normal vector
        field to $\Sigma$ (uniquely defined up to a sign in the
        timelike or spacelike case but not in the null case) normalized by
\begin{equation}\label{n}
        n\,.\,n\,=\,\epsilon\hspace{0.7cm},
\end{equation}
$\epsilon$ being positive, negative or zero respectively when the
hypersurface is timelike, spacelike or null.
It satisfies $n\, .\,e_{(a)}\,=\,0$. Let $\Phi(y^{\mu})\,=\,0$ be the
local equation of the hypersurface in this coordinate system and
assume that the normal is related to the gradient via
\begin{equation}
                n_{\mu}\,=\,\alpha^{-1}\,\partial_{\mu}\,\Phi\hspace{0.7cm},
\end{equation}
where $\alpha$ is a non-vanishing function of the local coordinates.
We introduce a transverse vector field \cite{BI} -see figure 1-
non uniquely defined by
\begin{equation}\label{N}
                N.n \,=\, \eta^{-1}\hspace{0.7cm},
\end{equation}
where $\eta$ is a non-vanishing function defined on $\Sigma$. For a
fixed $\eta$, $N$ is free up to a tangential displacement
\begin{equation}\label{gaugeN}
        N \rightarrow N+\lambda^{a}(\xi^b)\,e_{(a)}\hspace{0.7cm},
\end{equation}
where $\lambda^{a}$ is an arbitrary vector field on
$\Sigma$. According to the regularity condition (\ref{RC3}), we will
employ the following notation
\begin{equation}\label{NotJumps}
w^{\mu}_{\ \nu\rho} \,=\, \lbrack \Gamma^{\mu}_{\ \nu\rho} \rbrack
\hspace{0.7cm}.
\end{equation}
Because the metric is supposed to be only continuous, the Riemannian
part of the connection ${^R}\Gamma^{\lambda}_{\ \mu\nu}$ is also
jumping across $\Sigma$ and we get from (\ref{RelCon})
\begin{equation}\label{RelJumpCon}
        w^{\mu}_{\ \nu\rho} \,=\, {^R}w^{\mu}_{\ \nu\rho}+x^{\mu}_{\
        \nu\rho}
      \hspace{0.7cm},
\end{equation}
where ${^R}w^{\mu}_{\ \nu\rho}$ is the jump across $\Sigma$ of the
Riemannian connection ${^R}\Gamma^{\mu}_{\ \nu\rho}$.
We now assume that the Einstein-Cartan equations
(\ref{FE1})-(\ref{FE2}) apply in the sense of distributions on the
manifold $\cal M$, following the now classical techniques first
introduced by Lichnerowicz \cite{Lichne} and Taub \cite{Taub}.
From the regularity conditions
(\ref{RC1})-(\ref{RC3}) and the general expression of the curvature
tensor (\ref{Curv}), the curvature tensor contains a distributional
Dirac $\delta(\Phi)$ part $\hat{R}_{\mu\nu\rho\lambda}$ which is given by
\begin{equation}\label{ImpRiem}
    \alpha^{-1}\,\hat{R}_{\mu\nu\lambda\rho} \,=\,
    n_{\lambda}w_{\mu\nu\rho}-n_{\rho}w_{\mu\nu\lambda}\hspace{0.7cm}.
\end{equation}
Requiring that the distributional curvature tensor must satisfy the
identity (\ref{sym3}) and extracting the impulsive part of this
identity, we arrive with the help of (\ref{ImpRiem}) at
\begin{equation}\label{**}
                n_{\lambda} w_{\mu\nu\rho}-n_{\rho} w_{\mu\nu\lambda}=
                n_{\mu}w_{\lambda\rho\nu}-n_{\nu}w_{\lambda\rho\mu}+
                x_{\mu\nu\rho}\,n_{\lambda}+x_{\nu\mu\lambda}\,n_{\rho}+
                x_{\rho\lambda\nu}\,n_{\mu}+x_{\lambda\rho\mu}\,n_{\nu}
                \hspace{0.7cm}.
\end{equation}
Taking into account the relation (\ref{RelJumpCon}), equation
(\ref{**}) turns out to be an identity involving only the Riemannian
connection, namely
\begin{equation}\label{***}
         n_{\lambda} {^R}w_{\mu\nu\rho}-n_{\rho} {^R}w_{\mu\nu\lambda}=
         n_{\mu}{^R}w_{\lambda\rho\nu}-n_{\nu}{^R}w_{\lambda\rho\mu}
                \hspace{0.7cm}.
\end{equation}
Contracting this last identity by $N^{\rho}$ leads to the following
algebraic form of ${^R}w_{\mu\nu\lambda}$
\begin{equation}\label{Rw}
        {\eta}^{-1}\,{^R}w_{\mu\nu\lambda} \,=\,{1\over 2}\left(
          p^{R}_{\mu\nu}\,n_{\lambda}+p^{R}_{\lambda\mu}\,n_{\nu}
                                -p^{R}_{\lambda\nu}\,n_{\mu}
          \right)
          \hspace{0.7cm},
\end{equation}
where we have introduced the quantity
\begin{equation}
        p^{R}_{\mu\nu}\,=\,2\,{^R}w_{\mu\nu\rho}\,N^{\rho}\hspace{0.7cm}.
\end{equation}
The symmetry of the Riemannian connection applied to the identity
(\ref{Rw}) now leads directly to the expression for the jump of the
first derivatives of the metric
\begin{equation}
                \lbrack \partial_{\nu} \, g_{\lambda\mu} \rbrack \,=\,
                                \eta\,\gamma^{R}_{\lambda\mu}\,n_{\nu}
                                \hspace{0.7cm},
\end{equation}
with
\begin{equation}
 \gamma^{R}_{\lambda\mu}\,=\,N^{\nu}\,\lbrack \partial_{\nu}
 \,g_{\lambda\mu}\rbrack
 \hspace{0.7cm},
\end{equation}
which is exactly the same result obtained in \cite{BI} in general
relativity, with the same notation. From relation ({\ref{RelJumpCon}),
  we now finally arrive at the jump of the Einstein-Cartan connection
\begin{equation}
       w_{\mu\nu\lambda}\,=\,{\eta\over 2}\left(
              \gamma^{R}_{\mu\nu}\,n_{\lambda}+\gamma^{R}_{\lambda\mu}\,n_{\nu}
                                -\gamma^{R}_{\lambda\nu}\,n_{\mu} \right)
                        +x_{\mu\nu\lambda}\hspace{0.7cm}.
\end{equation}
The impulsive part of the Riemann tensor (\ref{ImpRiem}) is therefore
\begin{equation}\label{ImpRiem2}
    \alpha^{-1}\,\hat{R}_{\mu\nu\lambda\rho} \,=\,
        n_{\lambda}x_{\mu\nu\rho}-n_{\rho}x_{\mu\nu\lambda}
        +{\eta \over 2}\,\left( \gamma^{R}_{\mu\rho}\,n_{\nu}\,n_{\lambda}
                        -\gamma^{R}_{\nu\rho}\,n_{\mu}\,n_{\lambda}+
                        \gamma^{R}_{\nu\lambda}\,n_{\mu}\,n_{\rho}
                        -\gamma^{R}_{\mu\lambda}\,n_{\nu}\,n_{\rho}
        \right)
        \hspace{0.7cm}.
\end{equation}
By successive contractions, we deduce from (\ref{ImpRiem2}) the
expression for the impulsive part of the Einstein-tensor
\begin{equation}
        \alpha^{-1}\,\hat{G}_{\nu\rho} \,=\, x_{\mu\nu\rho}\,n^{\mu}
        -x^{\mu}_{\,\nu\mu}\,n_{\rho}
        +x^{\lambda}_{\,\mu\lambda}n^{\mu}\,g_{\nu\rho}
        +\alpha^{-1}\,\hat{G}^R_{\nu\rho}\hspace{0.7cm},
\end{equation}
which clearly splits into two parts, a pure Einstein-Cartan part
involving the jump in the defect tensor $x_{\mu\nu\rho}$, i.e. the
jump in the torsion, and an Einstein-Riemann part
$\hat{G}^R_{\nu\rho}$ corresponding to the impulsive part of the
Einstein tensor obtained in general relativity \cite{BI} and given by
\begin{equation}
        \alpha^{-1}\,\hat{G}^R_{\nu\rho} \,=\,
                {\eta\over 2}\left( \gamma^{R}_{\nu}\,n_{\rho}
                  +\gamma^{R}_{\rho}\,n_{\nu}
           -\gamma^{R}\,n_{\nu}\,n_{\rho}-\gamma_{R}^{\dagger}\,g_{\nu\rho}
          -\epsilon\,(\gamma^{R}_{\nu\rho}-\gamma^{R}\,g_{\nu\rho})
                \right)
                \hspace{0.7cm},
\end{equation}
where we have set, following \cite{BI},
\begin{equation}
        \gamma^{R}_{\nu}\,=\,\gamma^{R}_{\nu\mu}\,n^{\mu}\,\,,\,\,
        \gamma^{R}\,=\,g^{\mu\nu}\,\gamma^{R}_{\mu\nu}\,\,,\,\,
        \gamma_{R}^{\dagger}\,=\,\gamma^{R}_{\mu\nu}\,n^{\mu}\,n^{\nu}
        \hspace{0.7cm}.
\end{equation}
From the field equation (\ref{FE1}), the distributional stress-energy
tensor $T^{\nu\rho}$ of the hybrid space-time $\cal M$ exhibits a
Dirac part $\alpha\,\Sigma_{\nu\rho}$ given by
\begin{equation}
        8\pi \, \alpha\,\Sigma^{\nu\rho}\,=\,\hat{G}^{\nu\rho}\hspace{0.7cm},
\end{equation}
which we can express as
\begin{equation}\label{4SET}
\Sigma^{\nu\rho}\,=\,\Sigma^{\nu\rho}_{R}
                        +{\scriptstyle{1\over 8\pi}}\left(
                        x^{\mu\nu\rho}\,n_{\mu}-x^{\mu\nu}_{\ \ \mu}\,n^{\rho}
                        +x^{\lambda}_{\,\mu\lambda}n^{\mu}\,g^{\nu\rho}
                                \right)\hspace{0.7cm},
\end{equation}
where $ \Sigma^{\nu\rho}_{R}$ is the stress-energy tensor of the shell
obtained in general relativity in \cite{BI} given by
\begin{equation}
16\pi \eta^{-1}\, \Sigma^{\nu\rho}_{R}\,=\,\gamma_R^{\nu}\,n^{\rho}
                +\gamma_R^{\rho}\,n^{\nu}
                -\gamma_R\,n^{\nu}\,n^{\rho}-\gamma_R^{\dagger}\,g^{\nu\rho}
                -\epsilon\,(\gamma_R^{\nu\rho}-\gamma_R\,g^{\nu\rho})
                \hspace{0.7cm}.
\end{equation}
Using the relations (\ref{contortion}), (\ref{FE2}) and
(\ref{NotJumps}), we can obtain the following useful form of
$\Sigma^{\nu\rho}$ in terms of the spin tensor
\begin{equation}\label{4SETspin}
\Sigma^{\nu\rho}\,=\,\Sigma^{\nu\rho}_{R}
                        -{\scriptstyle{1\over 2}}\lbrack
                        \left(S^{\mu\nu\rho}+S^{\nu\rho\mu}+S^{\rho\nu\mu}
                                        \right)\,n_{\mu}
                        \rbrack\hspace{0.7cm}.
\end{equation}
In order to interpret the quantity (\ref{4SET}) as the stress-energy
tensor of the matter shell having the hypersurface $\Sigma$ as history
in spacetime, our regularity conditions (\ref{RC1})-(\ref{RC3}) must
ensure that this tensor is indeed purely tangential to $\Sigma$. As
$\Sigma^{\nu\rho}$ is not symmetric in general, this leads to the two
conditions
\begin{equation}
        \mathrm{(a)}\,\,\Sigma^{\nu\rho}\,n_{\nu}\,=0\,\,\,;\,\,\,
        \mathrm{(b)}\,\,\Sigma^{\nu\rho}\,n_{\rho}\,=0\hspace{0.7cm}.
\end{equation}
It is easy to verify that condition (a) is automatically satisfied by
the algebraic form (\ref{4SET}) and this can been considered with our
assumptions (${\cal C}_{\rho}\,=\,0$) as an algebraic consequence of
the twice contracted Bianchi identity $\nabla_{\nu}\,G^{\nu}_{\mu}\,=\,0$ by
extracting its $\delta'$-distributional part. Condition (b) is
equivalent to the following condition
\begin{equation}\label{Constraint}
        x_{\mu\nu\rho}\,n^{\mu}\,n^{\rho}
        +(x^{\mu}_{\,\rho\mu}\,n^{\rho})\,n_{\nu}
        -\epsilon\,x^{\mu}_{\,\nu\mu}\,=\,0\hspace{0.7cm},
\end{equation}
and from (\ref{4SETspin}), we get that this condition is equivalent to
the following requirement on the spin tensor
\begin{equation}\label{ConstraintAlt}
        \lbrack S^{\mu}_{\ \nu\rho}\,n_{\mu}\,n^{\nu} \rbrack \,=\,0
        \hspace{0.7cm}.
\end{equation}
We shall analyze in detail the intrinsic content of this constraint
and prove that it is satisfied.

        We first consider the case of a timelike or spacelike shell
        ($\epsilon\,\neq\,0$). In this situation, the transversal $N$
        can be chosen to be the normal $n$ and the projector operator
        on $\Sigma$ is defined by
\begin{equation}\label{Proj}
        g^{ab}\,e_{(a)}^{\mu}\,e_{(b)}^{\nu}\,=\,g^{\mu\nu}
        -\epsilon^{-1}n^{\mu}\,n^{\nu}
        \hspace{0.7cm},
\end{equation}
where $g^{ab}$ represents the inverse induced metric which exists in
this case. The constraint (\ref{Constraint}) is now reduced to the
intrinsic trace-free condition
\begin{equation}\label{TLintConst}
                g^{bc}x_{bac}\,=\,x^{b}_{\ ab}\,=\,0\hspace{0.7cm}.
\end{equation}
From the definition (\ref{contortion}) and field equation (\ref{FE2}),
this three-dimensional constraint turns to be
\begin{equation}
        \lbrack S^{b}_{\ a b}\rbrack \,=\, 0\hspace{0.7cm},
\end{equation}
which is only a part of the constraint obtained in \cite{AKP} but is
weaker because we do not assume continuity of the extrinsic curvature.

        For a lightlike hypersurface ($\epsilon\,=\,0$), the induced
        metric is degenerate and an inverse does not exist. Following
        the technique introduced in \cite{BI}, we introduce a
        pseudo-inverse symmetric tensor $g^{ab}_*$ not uniquely
        defined by the following relation
\begin{equation}
                g_{*}^{ac}\,g_{cb}\,=\,{\delta}^{a}_{b}-\eta\,l^{a}\,N_{b}
                \hspace{0.7cm},
\end{equation}
where $N_b\,=\,N\,.\,e_{(b)}$ and where the components $l^{a}$ are
such that in the oblique basis $(N, \,e_{(a)})$, the normal vector $n$
expands as
\begin{equation}
               n\: =\: \epsilon \: \eta \: N + l^{a}\: e_{(a)} \hspace{0.7cm}.
\end{equation}
The relation (\ref{Proj}) is now generalized to
\begin{equation}
g^{\mu \nu}\,=\,g_{*}^{ab}\, e_{(a)}^{\mu} \,e_{(b)}^{\nu}
                +2\eta \, l^{a} e_{(a)}^{(\mu}N^{\nu)}
                +{\eta}^{2}\epsilon\,N^{\mu}N^{\nu}\hspace{0.7cm}.
\end{equation}
It can easily be proved that the constraint (\ref{Constraint}) reduces
now to the two conditions
\begin{eqnarray}\label{LLintConst}
        x_{abc}\,l^b\,l^c\,=\,0 \hspace{0.7cm},\\
        l^a\,g_{*}^{bc}\,x_{abc}\,=\,0\hspace{0.7cm}.
\end{eqnarray}
As we pointed out, the regularity condition (\ref{RC2}) is equivalent
to $x_{abc}\,=\,0$. The constraints (\ref{TLintConst}) and
(\ref{LLintConst}) respectively in the timelike/spacelike and
lightlike cases are thereby automatically satisfied proving that the
assumption (\ref{RC2}) is sufficient. We shall prove now that it is necessary.

        In order to describe the intrinsic matter content of the shell
represented by its three-dimensional stress-energy tensor, we
introduce, following reference \cite{BI}, the ``transverse'' extrinsic
curvature ${\cal K}_{ab}$ of the hypersurface $\Sigma$ generalizing
the usual extrinsic curvature (whose utility was confined to the
timelike and spacelike shells) via
\begin{equation}\label{Kab}
     {\cal K}_{ab}\:=\: -N\: .\: {\nabla}_{e_{(b)}}e_{(a)}\hspace{0.7cm}.
\end{equation}
In the Einstein-Cartan theory, this quantity is non symmetric.
As in general relativity, ${\cal K}_{ab}$ is not a three-tensor under
changes of intrinsic coordinates ${\xi}^{a}$. Moreover, it is not
invariant under the transformation on
$N$ (\ref{gaugeN}) but transforms according to
\begin{equation}\label{TransK}
        {\cal K}_{ab} \rightarrow {\cal K}_{ab}-\lambda^c\,\Gamma_{c,ab}\,=\,
                {\cal K}_{ab}-\lambda^c\,{^R}\Gamma_{c,ab}
                -\lambda^c\,\chi_{cab} \hspace{0.7cm},
\end{equation}
where $\Gamma_{c,ab}$ represents the three-dimensional connection
induced on the submanifold $\Sigma$ by the connection
$\Gamma^{\mu}_{\ \nu\lambda}$ on $\cal M$.
Let us now consider the jump of ${\cal K}_{ab}$ across $\Sigma$ by
considering as in reference \cite{BI} the following quantity
\begin{equation}
        \gamma_{ab}\,=\,2\,\lbrack {\cal K}_{ab} \rbrack\hspace{0.7cm}.
\end{equation}
In order to make the formalism independent of the choice of the
transversal $N$, we must require $\gamma_{ab}$ to be gauge invariant
under (\ref{gaugeN}). In general relativity, the Riemann-Christoffel
symbols ${^R}\Gamma_{c,ab}$ of the three-dimensional manifold $\Sigma$
are constructed only out of tangential derivatives of the metric which
must be continuous across $\Sigma$. The jump of $\gamma_{ab}$ across
$\Sigma$ is therefore zero. This gauge invariance is thereby a direct
consequence of the regularity condition (\ref{RC1}). In the
Einstein-Cartan theory, the jump of the Riemann-Christoffel symbols
associated to the metric is also zero but the transformation
(\ref{gaugeN}) gives an additional term in (\ref{TransK}) which
involves the defect. It is now clear that the gauge invariance of
$\gamma_{ab}$ requires the jump $\lambda^c\,x_{cab}$ to be zero for an
arbitrary vector field $\lambda^c(\xi^a)$ on $\Sigma$, and thus
$x_{abc}\,=\,0$. This leads directly to the condition
(\ref{RC2}). This condition also ensures that, under change of
intrinsic coordinates $\xi^b\rightarrow \xi^{'a}(\xi^b)$, the quantity
$\gamma_{ab}$ behaves as a three tensor intrinsic to $\Sigma$.

        The four-dimensional tensor $\Sigma^{\mu\nu}$ (\ref{4SET}) of
the shell being purely tangential to $\Sigma$, there exists a
three-dimensional tensor $\Sigma^{ab}$ such that
\begin{equation}
        \Sigma^{\mu\nu}\,=\,\Sigma^{ab}\,e_{(a)}^{\mu}\,e_{(b)}^{\nu}
        \hspace{0.7cm},
\end{equation}
which is not simply obtained from
$\Sigma_{ab}\,=\,e_{(a)}^{\mu}\,e_{(b)}^{\nu}\,\Sigma_{\mu\nu}$
by raising the three-indices since in general, the induced metric
could be degenerate. After some algebra, we obtain the following
intrinsic stress-energy tensor of the shell
\begin{equation}\label{3SET}
     16\pi \, {\eta}^{-1}\,\Sigma^{ab}\,=\left(g_{*}^{ac}\,l^{b}l^{d}
      +l^{a}l^{c}\,g_{*}^{bd}-g_{*}^{ab}\,l^{c}l^{d}-l^{a}l^{b}\,g_{*}^{cd}
                                                        \right){\gamma}_{cd}
   -\epsilon \,\left(\,g_{*}^{ac}g_{*}^{bd}-g_{*}^{ab}g_{*}^{cd}
                                         \, \right){\gamma}_{cd}
                                       \hspace{0.7cm},
\end{equation}
which is formally the same expression as in general relativity but
$\gamma_{ab}$ is now a non symmetric quantity which splits into a
Riemann part and a Cartan part, namely
\begin{equation}\label{3gamma}
        \gamma_{ab}\,={^R}\gamma_{ab}+\beta_{ab}\hspace{0.7cm},
\end{equation}
where ${^R}\gamma_{ab}$ is the Riemannian quantity used in \cite{BI},
i.e. the jump in the Riemannian transverse extrinsic curvature defined
by the formula (\ref{Kab}) but calculated with the Riemannian
connection. In (\ref{3gamma}), we introduced the projection
$\beta_{ab}$ of the tensor $\beta_{\mu\nu}$ defined by
\begin{equation}\label{beta}
        \beta_{\mu\nu}\,=\,-2\,N^{\rho}\,x_{\rho\mu\nu} \hspace{0.7cm}.
\end{equation}
The quantity (\ref{3SET}) is, as in general relativity, both
independent of the choice of the transversal and of the choice of the
pseudo-tensor $g^{ab}_*$ and this independence is mainly ensured by
the continuity requirement $x_{abc}\,=\,0$. Let us insist on the fact
that, as in general relativity, despite the fact that the four
dimensional expression (\ref{4SET}) has been obtained using
distributional theory applied in a smooth coordinate system, the
construction of the intrinsic expression (\ref{3SET}) does not require
the construction of such spacetime coordinates that match continuously
at $\Sigma$ and in which the four dimensional metric is
continuous. The expression (\ref{3SET}) can be calculated on
either side of the shell using its adapted coordinates.

        Let us now summarize the procedure to construct the surface
stress-energy tensor of the shell :
\begin{trivlist}
        \item[(1)] In the hybrid spacetime ${\cal M}\,=\,M^+\cup M^-$
          where  $\Sigma$ is embedded, choose a transversal $N$ and a
          normal  $n$ such that (\ref{n}) and (\ref{N}) are satisfied
          and  such that
\begin{equation}
        \lbrack N_a \rbrack\,=\,\lbrack N\,.\, e_{(a)}\rbrack\,=\,0
        \hspace{0.7cm},
\end{equation}
         in order for $N$ to be geometrically well defined in $\cal M$
         as it  can be given in $M^+$ and $M^-$ by two different sets
         of  coordinates $\lbrace N_{\alpha}^+\rbrace$ and
         $\lbrace N_{\alpha}^-\rbrace$.
        \item[(2)] Choose one of the two coordinates systems
          $\{x^{\alpha}_-\}$ or $\{x^{\alpha}_+\}$ (we then drop the
          positive or negative suffix) and extend the quantity
          $\gamma_{ab}$ obtained from (\ref{3gamma}) to any four
          tensor $\gamma_{\mu\nu}$ such that
\begin{equation}
                \gamma_{\mu\nu}\,e_{(a)}^{\mu}\,e_{(b)}^{\nu}\,=\,\gamma_{ab}
                \hspace{0.7cm}.
\end{equation}
         Here $\gamma_{\mu\nu}$ has the form
         $\gamma_{\mu\nu}\,=\,{^R}\gamma_{\mu\nu}+\beta_{\mu\nu}$
         where ${^R}\gamma_{\mu\nu}$ is any four tensor such that
     ${^R}\gamma_{\mu\nu}\,e_{(a)}^{\mu}\,e_{(b)}^{\nu}\,=\,{^R}\gamma_{ab}$.
           Thus we obtain the four dimensional form (\ref{4SET}). We can
        equivalently obtain directly the intrinsic form
          (\ref{3SET}) by means of the pseudo-inverse $g_{*}^{ab}$.
\end{trivlist}

        In the case of a lightlike shell, the intrinsic stress-energy
        tensor  (\ref{3SET}) is reduced to
\begin{equation}
        16\pi \, {\eta}^{-1}\,\Sigma^{ab}\,=\,
                g_{*}^{ac}\,l^{b}\left(\gamma_{cd}\,l^d\right)
                +l^a g_{*}^{bd}\left(l^c\,\gamma_{cd}\right)
                -g_{*}^{ab}\left(\gamma_{cd}\,l^c\,l^d\right)
                -l^{a}l^{b}\left(g_{*}^{cd}\,\gamma_{cd}\right)
                \hspace{0.7cm},
\end{equation}
showing, as in general relativity -see reference \cite{BBH1}, that
there is a part
$\hat{\gamma}_{ab}$ of  $\gamma_{ab}$ which does not contribute to
$\Sigma^{ab}$.
$\hat{\gamma}_{ab}$ satisfies the following system of 7 equations
\begin{equation}\label{S}
    \left\lbrace
        \begin{array}{ccc}
                \hat{\gamma}_{cd}\,l^d & = & 0 \\
                \hat{\gamma}_{cd}\,l^c & = & 0 \\
                g_{*}^{cd}\,\hat{\gamma}_{cd}& = & 0
        \end{array}
    \right. \hspace{0.7cm},
\end{equation}
and is the generalisation of the quantity obtained in reference
\cite{BBH1} (where $\gamma_{ab}$ was symmetric and therefore
$\hat{\gamma}_{ab}$ satisfied only 4 equations). From the general
splitting (\ref{3gamma}), it is clear that
$\hat{\gamma}_{ab}$ is obtained as
\begin{equation}
        \hat{\gamma}_{ab}\,=\,{^R}\hat{\gamma}_{ab}+\hat{\beta}_{ab}
        \hspace{0.7cm},
\end{equation}
where ${^R}\hat{\gamma}_{ab}$ is the symmetric quantity obtained in
\cite{BBH1} and where $\hat{\beta}_{ab}$ is the part of $\beta_{ab}$
which satisfies the system (\ref{S}). If we choose the transversal to
be a null vector (which is always possible), one obtains
\begin{equation}\label{hatgam}
        \hat{\gamma}_{ab}\,=\,\gamma_{ab}
        -{\scriptstyle{1\over 2}}\left(g_{*}^{cd}\,\gamma_{cd}\right)g_{ab}
                -\eta\,\left( l^c\,\gamma_{cb}\right)N_a
                -\eta\,\left( \gamma_{ac}\,l^c\right)N_b
                +\eta^2\,\left(\gamma_{cd}\,l^c\,l^d\right)N_a N_b
                \hspace{0.7cm},
\end{equation}
with
\begin{equation}
        \hat{\beta}_{ab}\,=\,\beta_{ab}
        -{\scriptstyle{1\over 2}}\left(g_{*}^{cd}\,\beta_{cd}\right)g_{ab}
                -\eta\,\left( l^c\,\beta_{cb}\right)N_a
                -\eta\,\left( \beta_{ac}\,l^c\right)N_b
                +\eta^2\,\left(\beta_{cd}\,l^c\,l^d\right)N_a N_b
                \hspace{0.7cm},
\end{equation}
and with \cite{BBH1}
\begin{equation}
        {^R}\hat{\gamma}_{ab}\,=\,{^R}\gamma_{ab}
        -{\scriptstyle{1\over 2}}\left(g_{*}^{cd}\,{^R}\gamma_{cd}\right)g_{ab}
                -2\eta\,l^c\,{^R}\gamma_{c(a}\,N_{b)}
                +\eta^2\,\left({^R}\gamma_{cd}\,l^c\,l^d\right)N_a N_b
                \hspace{0.7cm}.
\end{equation}

By analogy with general relativity, we interpret $\hat{\gamma}_{ab}$
(the part of $\gamma_{ab}$ which does not contribute to the expression
of the intrinsic stress-energy tensor $\Sigma^{ab}$ of the null shell)
as being an impulsive gravitational wave propagating in $\cal M$ whose
history in $\cal M$ is the null hypersurface $\Sigma$ and which
propagates independently of the null shell. Since $\gamma_{ab}$ has 9
independent components and as $\hat{\gamma}_{ab}$ satisfies the system
(\ref{S}) of 7 independent equations, it follows that
$\hat{\gamma}_{ab}$ has two independent components which can be
interpreted as representing the two degrees of polarization of the
gravitational impulsive wave.

        In the following section, we will apply this formalism to
the construction of a null shell.


\section{Application: an example of a null shell}\label{IV}

        We first construct a solution of the Einstein-Cartan
equations which has an easily identifiable family of null hypersurfaces. To
achieve this we look for a metric tensor of the general static form
\begin{equation}\label{ds}
        ds^2\,=\,-du\,\left( f(r)\,du+2\,dr\right)+r^2\left(
                                d\theta^2+\sin^2\theta\,d\phi^2\right)
                              \hspace{0.7cm},
\end{equation}
and a torsion tensor with all components vanishing except
\begin{equation}\label{Qform}
        Q^{u}_{\ u r}\,=\,\alpha(r)\hspace{0.7cm}.
\end{equation}

In order to apply the shell formalism of the preceeding section, we
first impose the space-time constraint (\ref{STC}) which corresponds to
the vanishing of the components of the vector field ${\cal
  C}_\rho$. These components vanish identically except for
\begin{equation}
  {\cal C}_u\,=\,\frac{\alpha\,\left( f'+2\,\alpha\,f\right)}{r}
  \hspace{0.7cm}.
\end{equation}
This constraint leads therefore immediately to
\begin{equation}\label{alpha}
        \alpha(r)\,=\,-{1\over 2}\,\frac{f'}{f}\hspace{0.7cm},
\end{equation}
where the $'$ stands for $d/dr$.

The evolution equation for the torsion tensor is
\begin{equation}
        \nabla_{\mu}\,Q^{\mu}_{\ \nu\rho}\,=\,2\,G_{[\nu\rho]}
        -Q^{\mu}_{\ \kappa\mu}\,Q^{\kappa}_{\ \nu\rho}\hspace{0.7cm}.
\end{equation}
But with the metric (\ref{ds}) and the torsion (\ref{Qform}), the
Einstein tensor turns out to be symmetric and with (\ref{alpha}) is
given by
\begin{equation}
  \label{eq:Gmunu}
  G_{\mu\nu}\,=\,\left [
    \begin {array}{cccc}
      -{\frac {f\left (f'\,r-1+f \right)}{{r}^{2}}}&
      -{\frac {f'\,r-1+f}{{r}^{2}}}&0&0\\
  \noalign{\medskip}-{\frac {f'\,r-1+f}{{r}^{2}}}&
      -{\frac {f'}{rf}}&0&0\\
  \noalign{\medskip}0&0&{\frac{1}{2}}\,r\,f'&0\\
  \noalign{\medskip}0&0&0&{\frac{1}{2}}\,r\,f'\,\sin^2\theta
      \end {array}
    \right ]\hspace{0.7cm}.
\end{equation}
It can be checked that our assumptions lead to
\begin{equation}
  Q^{\mu}_{\ \kappa\mu}\,Q^{\kappa}_{\ \nu\rho}\,=\,0\hspace{0.7cm}.
\end{equation}
As a result, the evolution equation for the torsion tensor reduces
to
\begin{equation}
  \nabla_{\mu}\,Q^{\mu}_{\ \nu\rho}\,=\,0\hspace{0.7cm},
\end{equation}
an equation which is now identically satisfied. It can also be
checked after tedious algebra that the two Bianchi identities
(\ref{cyclic}) and (\ref{Bianchi}) are both satisfied by virtue of the
torsion choice (\ref{alpha}). The stress-energy tensor of the
spacetime is therefore given by
$T_{\mu\nu}\,=\,1/8\pi\,G_{\mu\nu}$. This $T_{\mu\nu}$
is a conserved symmetric tensor whose proper values are
all real \footnote{These proper values $\lambda$, obtained as the roots of the
  equation $\mathrm{det}(T_{\alpha\beta}-\lambda\,g_{\alpha\beta})\,=\,0$, are
$\lambda_0\,=\,{rf'+f-1\over 8\pi\,r^2}$,
$\lambda_1\,=\,{f-1\over 8\pi\,r^2}$ and
$\lambda_2\,=\,{f'\over 16\pi\,r}$ (double root).}.

From the second Cartan field equation (\ref{FE2}), we obtain the
only non zero components of the spin tensor:
\begin{equation}
  S^{\theta}_{\ r\theta}\,=\,S^{\phi}_{\ r\phi}\,=\,
  -{\alpha\over 8\pi}\,=\,{1\over 16\pi}\,\frac{f'}{f}\hspace{0.7cm}.
\end{equation}

In the spacetime with metric (\ref{ds}), the $u\,=\,const$ form a
family of null hypersurfaces. Using the technique described in the
preceeding section, we would like to glue two spacetimes $M_+$
and $M_-$ both endowed with a metric of the general form
(\ref{ds}) in two different coordinate
systems $(u,r_+,\theta_+,\phi_+)$ and $(u,r,\theta,\phi)$
with torsion given by (\ref{Qform}) and (\ref{alpha}) but
with different functions $f_+(r_+)$ and $f_-(r)$ and different
$\alpha_+(r_+)$ and $\alpha_-(r)$. We join these spacetimes along
the null hypersurface $\Sigma$ with equation $u\,=\,0$ by
requiring the positive and negative sides of $\Sigma$ to be isometrically
soldered via the identity matching
\begin{equation}
  \label{eq:Id}
  r_+\,=\,r\,\,,\,\,\theta_+\,=\,\theta\,\,,\,\,\phi_+\,=\,\phi
  \hspace{0.7cm}.
\end{equation}
This enables the metric continuity condition (\ref{RC1}) to be
automatically satisfied, the common induced metric being
$ds^2_{\vert_{\Sigma}}\,=\,r^2\,d\Omega^2$ and $r$ is chosen as a
common parameter along the generators of each side of the
hypersurface. Let us choose $\xi^a\,=\,(r,\theta.\phi)$ as the
intrinsic coordinates on the three-dimensional manifold $\Sigma$.
The continuity condition (\ref{RC2}) for the torsion
tensor is then automatically satisfied since on both sides of $\Sigma$,
the torsion tensor satisfies $Q_{abc}\,=\,0$ as well as the necessary
condition $\lbrack S_{abc}+S_{bca}+S_{cab}\rbrack\,=\,0$.

The future-directed normal generator of $\Sigma$ is the null vector
\begin{equation}
  n\,=\,{\partial \over \partial r}\hspace{0.7cm},
\end{equation}
and as transversal we choose the null vector
\begin{equation}\label{N2}
  N\,=\,{\partial \over \partial u}
         -{\scriptstyle{1\over 2}}\,f\,{\partial \over \partial r}
         \hspace{0.7cm}.
\end{equation}
On both sides of $\Sigma$, we have
\begin{eqnarray}
  N\,.\,N&=&0\hspace{0.7cm},\\
  N\,.\,n&=&\eta\,=\,-1\hspace{0.7cm},\\
  N\,.\,e_{(A)}&=&0,\,\,(A=2,3)\hspace{0.7cm},
\end{eqnarray}
which makes $N$ a geometrically well-defined object.

The non zero components of the Riemannian transverse extrinsic
curvature are
\begin{eqnarray}
  \label{eq:KabR}
  K^{R}_{\theta\theta}\vert_{\pm}&=&-{\scriptstyle{1\over2}}\,r\,f^{\pm}
  \hspace{0.7cm},\\
  K^{R}_{\phi\phi}\vert_{\pm}&=&
                    -{\scriptstyle{1\over2}}\,r\,\sin^2\theta\,f^{\pm}
                    \hspace{0.7cm},
\end{eqnarray}
from which we get that the Riemannian part $\gamma^{R}_{ab}$ of
$\gamma_{ab}$ is
\begin{equation}
  \label{eq:gamR}
  \gamma^{R}_{ab}\,=\,\left[
  \begin{array}{ccc}
    0&0&0\\
    0&-r\,\left[ f \right]&0\\
    0&0&-r\,\sin^2\theta\,\left[ f \right]
  \end{array}
  \right]\hspace{0.7cm}.
\end{equation}
Since $l^a\,=\,\delta^a_r$, $N_b\,=\,N\,.\,e_{(b)}\,=\,-\delta^r_b$
and $\eta\,=\,-1$, we can choose $g^{ab}_*$ as being $g^{AB}$
($A,B=2,3$) bordered by zeros and take
\begin{equation}
  \label{eq:g*ab}
  g_*^{ab}\,=\,\left[
  \begin{array}{ccc}
    0&0&0\\
    0&r^{-2}&0\\
    0&0&r^{-2}\,\sin^{-2}\theta
  \end{array}
  \right]\hspace{0.7cm},
\end{equation}
and according to (\ref{3SET}) and (\ref{3gamma}), the Riemann part of
the stress-energy tensor of the shell is given by
\begin{equation}
  \label{eq:setRab}
  16\pi\,\Sigma^{ab}_R\,=\,-2\,\frac{\left[
  f\right]}{r}\,l^a\,l^b\,=\,4\pi\,\frac{\left[m\right]}{r^2}\,l^a\,l^b
     \hspace{0.7cm},
\end{equation}
where we introduced as in \cite{BI} a local mass function $m(r)$
given by $f(r)\,=1-2\,m/r$.

In order to determine the Cartan part of the surface stress-energy
tensor, we now calculate the tensor $\beta_{ab}$ defined in
(\ref{beta}). With $N$ given by (\ref{N2}), it turns to be
$\beta_{ab}\,=\,-2\,x_{uab}$. It is not difficult to see that $x_{urr}$
is the only non zero component of $x_{uab}$ and we get
\begin{equation}
  \label{eq:betaComp}
  \beta_{ab}\,=\,2\,\left[ \alpha\right]\,\delta^r_a\,\delta^r_b
  \hspace{0.7cm}.
\end{equation}
The Cartan part of the surface stress-energy tensor is finally
obtained as
\begin{equation}
  \label{eq:setCab}
  8\pi\,\Sigma^{ab}_C\,=\,\left[ \alpha \right]\,g^{ab}_*\hspace{0.7cm}.
\end{equation}
In conclusion, we have found that the hypersurface $\Sigma$ is the
history of a null shell of matter whose surface stress-energy tensor
is given by
\begin{equation}
  \label{eq:setab}
  8\pi\,\Sigma^{ab}\,=\,2\,\frac{\left[m\right]}{r^2}\,l^a\,l^b
          +\left[ \alpha \right]\,g^{ab}_*\hspace{0.7cm}.
\end{equation}
This means that the matter on the shell is characterized by a surface
energy density $\sigma$ expressed as
\begin{equation}
  \label{eq:sigma}
  4\pi\,r^2\,\sigma\,=\,-\left[ m \right]\hspace{0.7cm},
\end{equation}
which owes its existence to the jump in the mass function $m$, and
by a pressure $P$ given by
\begin{equation}
  \label{eq:p}
  8\pi\,P\,=\,\left[ \alpha \right]\hspace{0.7cm},
\end{equation}
whose existence is due to the jump in the torsion function. Finally,
it can be easily seen that the wave part $\hat{\gamma}_{ab}$ defined in
(\ref{hatgam}) vanishes identically showing that there is no
gravitational impulsive wave associated with this null shell of matter.

\section*{Acknowledgements}
I would like to express my gratitude to Professor
P. A. Hogan for many valuable and stimulating discussions, comments
and for constant advice and encouragement. The author wishes also to
thank the Departement of Education and Science (HEA) for a
post-doctoral fellowship.

\end{document}